\documentstyle[twocolumn,aps,prl,array,epsf]{revtex}
\begin{document}
\draft
\title{Quantum key distribution without alternative measurements\thanks{
Phys. Rev. A {\bf 61}, 052312 (2000).
After its publication, Zhang, Li, and Guo showed that
the protocol is insecure against a particular eavesdropping attack
(quant-ph/0009042).
A modified version which avoids this attack is presented
in quant-ph/0009051.}}
\author{Ad\'{a}n Cabello\thanks{Electronic address:
adan@cica.es, fite1z1@sis.ucm.es}}
\address{Departamento de F\'{\i}sica Aplicada,
Universidad de Sevilla, 41012 Sevilla, Spain}
\date{\today}
%First version: 28 June 1999.
%This version: 21 March 2000.
%Footnote: 2 October 2000.
%Esta versi\'{o}n incluye los cambios hechos en las pruebas de imprenta de PRA.
%y referencias al comment de Zhang, Li, y Guo, y a la vesi\'on modificada.
\maketitle
\begin{abstract}
Entanglement swapping between Einstein-Podolsky-Rosen (EPR) pairs
can be used to generate
the same sequence of random bits in two remote places. A quantum
key distribution protocol based on this idea is described. The
scheme exhibits the following features.
(a) It does not require that
Alice and Bob choose between alternative measurements,
therefore improving the rate
of generated bits by transmitted qubit. (b) It allows Alice and
Bob to generate a key of arbitrary length using a single quantum
system (three EPR pairs), instead of a long sequence of them.
(c) Detecting Eve requires the
comparison of fewer bits. (d) Entanglement is an essential
ingredient. The scheme assumes reliable measurements of the Bell
operator.
\end{abstract}

\narrowtext
The two main goals of cryptography are for two distant
parties, Alice and Bob, to be able to communicate in a form that
is unintelligible to a third party, Eve, and to prove that the
message was not altered in transit. Both of these goals can be
accomplished securely if both Alice and Bob are in possession of
the same secret random sequence of bits, a ``key''
\cite{Vernam}. Therefore, one of the main problems of cryptography
is the key distribution problem, that is, how do Alice and Bob,
who initially share no secret information, come into the possession of
a secret key, while being sure that Eve cannot acquire even
partial information about it. This problem cannot be solved by
classical means, but it can be solved using quantum mechanics
\cite{BB84}. The security of protocols for quantum key
distribution (QKD) such as the Bennett-Brassard 1984
(BB84) \cite{BB84}, E91 \cite{Ekert91},
B92 \cite{B92}, and other protocols \cite{GV95,ABCL98}, is assured by the
fact that while information stored in classical form can be
examined and copied without altering it in any detectable way, it
is impossible to do that when information is stored in unknown
quantum states, because an unknown quantum state cannot be
reliably cloned (``no-cloning'' theorem \cite{WZ82}). In these
protocols security is assured by the fact that both Alice and Bob
must choose randomly between two possible measurements.
In this paper I introduce a QKD scheme
which does not
require that Alice and Bob choose between alternative measurements.
This scheme is based on
``entanglement swapping'' \cite{telep1,es2,Cohen} between two
pairs of ``qubits'' (quantum two-level systems), induced by a Bell
operator measurement \cite{BMR92}.
The Bell operator is a
nondegenerate operator which acts on a pair of qubits $i$ and $j$,
and projects their combined state onto one of the four Bell
states
\begin{eqnarray}
\left| {00} \right\rangle _{ij} & = &
{1 \over {\sqrt 2}}
\left( {\left| 0 \right\rangle _i\otimes \left| 0
\right\rangle _j +
\left| 1 \right\rangle _i\otimes \left| 1
\right\rangle _j} \right), \\
\left| {01} \right\rangle _{ij} & = &
{1 \over {\sqrt 2}}
\left( {\left| 0 \right\rangle _i\otimes \left| 0
\right\rangle _j -
\left| 1 \right\rangle _i\otimes \left| 1
\right\rangle _j} \right), \\
\left| {10} \right\rangle_{ij} & = &
{1 \over {\sqrt 2}}
\left( {\left| 0 \right\rangle _i\otimes \left| 1
\right\rangle _j +
\left| 1 \right\rangle _i\otimes \left| 0
\right\rangle _j} \right), \\
\left| {11} \right\rangle_{ij} & = &
{1 \over {\sqrt 2}}
\left( {\left| 0 \right\rangle _i\otimes \left| 1
\right\rangle _j -
\left| 1 \right\rangle _i\otimes \left| 0
\right\rangle _j} \right).
\end{eqnarray}
Entanglement swapping works as follows. Consider a pair of qubits,
$i$ and $j$, prepared in one of the four Bell states, for
instance,
$\left| {11} \right\rangle_{ij}$. Consider a second
pair of qubits $k$ and $l$ prepared in another Bell state, for
instance, $\left| {01} \right\rangle_{kl}$. If a Bell operator
measurement is performed on $i$ and $k$, then the four
possible results ``$00$,'' ``$01$,'' ``$10$,'' and ``$11$'' have
the same probability to occur. In fact, the outcome of each
measurement is purely random. Suppose that the result ``$00$'' is
obtained, consequently the state of the pair $i$ and $k$ after the
measurement is $\left| {00} \right\rangle_{ik}$. Moreover, the
state of $j$ and $l$ is projected onto state
$\left| {10} \right\rangle_{jl}$. Therefore, the state of
$j$ and $l$ becomes entangled although they have never interacted.

I will denote the initial state of the pairs $i$, $j$ and $k$,
$l$, in the previous example by $\left| {11}
\right\rangle_{ij}\otimes\left| {01} \right\rangle_{kl}$, and the
final state of the pairs $i$, $k$ and $j$, $l$ by $\left| {00}
\right\rangle_{ik}\otimes\left| {10} \right\rangle_{jl}$. Suppose
that the initial state of the pairs $i$, $j$ and $k$, $l$ is a
product of two Bell states and, as in the previous example, a Bell
operator measurement is executed on two qubits, one of each pair;
then, after the measurement the state of the pairs $i$, $k$ and
$j$, $l$ becomes a product of two Bell states. All possibilities
are collected in Table~I.

The proposed scheme for QKD is illustrated in Fig. 1 and it is
described as follows.

(i) Consider six qubits numbered $1$ to $6$. Alice prepares
qubits $1$ and $2$ in the Bell state $\left| {11}
\right\rangle_{12}$, and qubits $3$ and $5$ in the Bell state
$\left| {10} \right\rangle_{35}$. In a remote place, Bob prepares
qubits $4$ and $6$ in the Bell state $\left| {10}
\right\rangle_{46}$. All this information is public. $2$
and $6$ will be the only transmitted qubits during the process.
Alice will always retain qubits $1$, $3$, and $5$; and Bob will
always retain qubit $4$.

(ii) Alice transmits qubit $2$ to Bob using a public
channel. This channel must be a transmission medium that isolates
the state of the qubit from interactions with the environment.

(iii) Alice secretly measures the Bell operator on qubits
$1$ and $3$, and Bob secretly measures the Bell operator on qubits
$2$ and $4$. The results of both experiments are correlated,
although Alice and Bob do not know how as yet. The purpose of the
next step is to elucidate how the results are correlated without
publicly revealing either of them.

(iv) Bob transmits qubit $6$ to Alice using a public
channel. Then Alice measures the Bell operator on qubits $5$ and
$6$, and publicly announces the result. Suppose that Alice has
obtained ``$11$'' in her secret measurement on qubits $1$ and $3$.
Then, since the initial state of $1$, $2$, $3$, and $5$ was
$\left| {11} \right\rangle_{12}\otimes\left| {10}
\right\rangle_{35}$, by using Table~I Alice knows that
the state of $2$ and $5$ is $\left| {10}
\right\rangle_{25}$. In addition, suppose that Alice obtains
``$00$'' in the public measurement on $5$ and $6$. Then,
since she knows that the previous state of $2$, $4$, $5$,
and $6$ was $\left| {10} \right\rangle_{25}\otimes\left| {10}
\right\rangle_{46}$, by using Table~I Alice knows that
Bob has obtained ``$00$'' in his secret measurement on $2$
and $4$. Following a similar reasoning, Bob can know that Alice
has obtained ``$11$'' in her secret measurement on $1$ and
$3$. Previously, Alice and Bob have agreed to choose the sequence
of results of Alice's secret measurements to form the key. The two
initial bits of the key are therefore ``$11$.'' The public
information shared by Alice and Bob is not enough for Eve to
acquire any knowledge of the result obtained by one of the parts.
Using this information Eve only knows that one of the following
four possible combinations of results for Alice and Bob's secret
measurements have occurred: ``$00$'' for Alice's result and
``$11$'' for Bob's, ``$01$'' and ``$10$,'' ``$10$'' and ``$01$,''
and ``$11$'' and ``$00$.''

One Bell state can be transformed into another just by rotating one
of the qubits. Using this property, Alice (Bob) can change the
Bell state of qubits $1$ and $3$ ($2$ and $4$)
to a previously agreed public state. Then
the situation is similar to (i)
and the next stage of the process can be started.

This scheme for QKD has the following features.

(a) It improves the rate of generated bits by transmitted
qubit. In BB84 and in B92 (and in E91), Bob (and Alice) must
choose between two alternative measurements in order to preserve
security. This implies that the number of useful random bits
shared by Alice and Bob by transmitted qubit, before checking for
eavesdropping, is $0.5$ bits by transmitted qubit, both in BB84
and B92 (and $0.25$ in E91), or at the most, it can be made to
approach 1 in Ref.~\cite{ABCL98}. In our scheme the rate is $1$ bit by
transmitted qubit. This is so because Alice and Bob always perform
the same kind of measurement, a Bell operator measurement, and
therefore, each of them acquires two correlated random bits after
each stage of the process. In each of these stages, only two
qubits are transmitted (one from Alice to Bob and another
from Bob to Alice). This improvement is very useful since a
key must be as large as the message to be transmitted (written as
a sequence of bits), and cannot be reused for subsequent messages
\cite{Vernam}.

(b) It only requires a single quantum system (three EPR pairs)
instead of a long sequence of quantum systems,
to generate a key of arbitrary length.
By contrast with previous schemes, in the
one presented here no source of qubits is needed.
The same two qubits (qubits 2 and 6) are transmitted to and
from Alice and Bob over and over again \cite{PRL1}.

(c) The detection of Eve requires the comparison of fewer
bits. The transmitted qubits do not encode the
bits that form the key, but only the type of correlation between
the results of the experiments that allow Alice and Bob to
secretly generate the key. Therefore, intercepting and copying
them does not allow Eve to acquire any information about the key. In
fact, the state of the transmitted qubits is public.
However, Eve can use a strategy ---also based on entanglement
swapping--- to learn Alice's sequence of secret results. This
strategy is illustrated in Fig.~2 and is described as follows.

(1a) Consider the same scenario as in (i) but
suppose Eve has two additional
qubits $7$ and $8$, initially prepared in a Bell state,
for instance, $\left| {00} \right\rangle_{78}$.

(1b) Eve intercepts qubit $2$ that Alice send to Bob
and makes a Bell operator measurement on qubits $2$ and $8$.
Then qubits $1$ and $7$ become entangled in a known (to Eve) Bell
state. For instance, if after Eve's measurement the state of $2$
and $8$ is $\left| {00} \right\rangle_{28}$, then the
state of $1$ and $7$ becomes $\left| {11} \right\rangle_{17}$.

(2) Therefore, after Eve's intervention the real situation
is not that described in (ii). Now qubit $1$ is
entangled with Eve's qubit $7$, and $2$ is entangled with
Eve's $8$.

(3a) In this new scenario, after Alice's (Bob's)
measurement on
qubits $1$ and $3$ ($2$ and $4$), the state of qubits $5$ and $7$
($6$ and $8$) becomes a Bell state. For instance, if Alice (Bob)
obtains ``$11$'' (``$00$''), the state of qubits $5$ and $7$ ($6$ and
$8$) would be $\left| {10} \right\rangle_{57}$ ($\left| {10}
\right\rangle_{68}$). However, these states are unknown to Eve,
because she (still) does not know the results of Alice's and Bob's
measurements.

(3b) Eve intercepts qubit $6$ that Bob sends to Alice
and makes a Bell operator measurement on qubits $6$ and $8$.
This reveals the state they were in. Then Eve can know Bob's
result. For instance, in our example, Eve would find ``$10$'' and
would know that Bob's result was ``$00$.''

(3c) Eve makes a Bell operator measurement on qubits $7$ and
$8$. Then qubits $5$ and $6$ becomes entangled in a Bell state
(still) unknown to Eve, because she does not know Alice's secret
result. For instance, if Eve obtains ``$01$,'' then qubits $5$ and
$6$ would be in the state $\left| {01} \right\rangle_{56}$.

(4) Eve gives qubit $6$ to Alice. Alice makes a measurement
on $5$ and $6$ and announces the result. Then Eve can
know the previous state of $5$ and $7$ ($\left| {10}
\right\rangle_{57}$, in our example) and the result of Alice's
measurement on $1$ and $3$ (``$11$,'' in our example).

However, Eve's intervention changes the correlation that Alice and
Bob expect between their secret results. For instance, in our
example, Bob, using his result and the result publicly announced
by Alice, thinks that the two initial bits of the key are
``$10$.''

As in previous QKD protocols, in our scheme Alice and Bob
can detect Eve's intervention by publicly comparing a sufficiently
large random subset of their sequences of bits, which they
subsequently discard. If they find that the tested subset is
identical, they can infer that the remaining untested subset is
also identical, and therefore can form a key. In BB84, for each
bit tested by Alice and Bob, the probability of that test
revealing the presence of Eve (given that Eve is indeed present)
is $\frac{1}{4}$. Thus, if $N$ bits are tested, the probability of
detecting Eve (given that she is present) is
$1-\left(\frac{3}{4}\right)^N$. In our scheme if Alice and Bob
compare a {\em pair} of bits generated in the same step, the
probability for that test to reveal Eve is $\frac{3}{4}$. Thus if
$n$ pairs ($N=2n$ bits) are tested, the probability of Eve's
detection is $1-\left(\frac{1}{2}\right)^N$.
This improvement in the efficiency of the detection of eavesdropping
has been pointed out for a particular eavesdropping
attack, it would be interesting to investigate
whether more general attacks exist and
whether the improvement in efficiency is also present in these cases.

(d) It uses entanglement as an essential tool. QKD was the
first practical application of quantum entanglement
\cite{Ekert91}. However, as shown in Ref.~\cite{BBM92}, entanglement
was not an essential ingredient, in the sense that almost the same
goals can be achieved without entanglement. However, subsequent
striking applications of quantum mechanics such as quantum dense
coding \cite{qdc1,qdc2}, teleportation of quantum states
\cite{telep1,telep2,telep3}, entanglement swapping
\cite{telep1,es2}, and quantum computation
\cite{qc1}, are strongly based on quantum entanglement.
The scheme described here relies on entanglement in the sense that
it performs a
task ---QKD with properties (a),
(b), and (c)---
that cannot be accessible without entanglement.

The practical feasibility of the scheme described in this paper
hinges on the feasibility of a reliable (i.e., with 100\%
theoretical probability of success) Bell operator measurement.
Bell operator measurements are also required for reliable double
density quantum coding and teleportation.
As far as I know,
the first proposals for a reliable Bell operator measurement
are those
which discriminate between the four polarization-entangled
two-photon Bell states using entanglement in additional degrees
of freedom \cite{KW98} or using atomic coherence \cite{SEB99}.

It is not expected that the protocol for QKD introduced in this
paper will be able to improve existing experiments \cite{exp} for real
quantum cryptography in practice.
Its main importance is conceptual: it
provides a different quantum solution to a
problem already solved by quantum mechanics.\\

The author thanks J. L. Cereceda, O. Cohen,
A. K. Ekert, C. Fuchs, T. Mor, and
B. Orfila for helpful comments.
This work was supported by the
Universidad de Sevilla (Grant No.~OGICYT-191-97) and the Junta de
Andaluc\'{\i}a (Grant No.~FQM-239).\\

\newpage

\begin{table}
\begin{center}
\begin{tabular}{cccccccc}
\hline
\hline
\multicolumn{4}{c}
{{\small Initial state $\left| {ijkl} \right\rangle$}} &
\multicolumn{4}{c}
{{\small Possible final states $\left| {ikjl} \right\rangle$}} \\ \hline
$\;0000\,$ & $\,0101\,$ & $\,1010\,$ & $\,1111\;$ &
$\;0000\,$ & $\,0101\,$ & $\,1010\,$ & $\,1111\;$ \\ \hline
$\;0001\,$ & $\,0100\,$ & $\,1011\,$ & $\,1110\;$ &
$\;0001\,$ & $\,0100\,$ & $\,1011\,$ & $\,1110\;$ \\ \hline
$\;0010\,$ & $\,0111\,$ & $\,1000\,$ & $\,1101\;$ &
$\;0010\,$ & $\,0111\,$ & $\,1000\,$ & $\,1101\;$ \\ \hline
$\;0011\,$ & $\,0110\,$ & $\,1001\,$ & $\,1100\;$ &
$\;0011\,$ & $\,0110\,$ & $\,1001\,$ & $\,1100\;$ \\ \hline
\hline
\end{tabular}
\end{center}
\vspace{0.2cm}
\noindent TABLE I.
{\small All possible results of a Bell operator measurement
on qubits $i$ and $k$. For example, if the initial state is
$\left| {11} \right\rangle_{ij}\otimes\left| {01} \right\rangle_{kl}$,
you must locate $1101$ on the left half of the table.
Then, after a Bell operator measurement
on $i$ and $k$, the four possible final states are represented
on the right half of the table by $0010$, $0111$, $1000$, and $1101$; where,
for instance, $0010$ means
$\left| {00} \right\rangle_{ik}\otimes\left| {10} \right\rangle_{jl}$.}
%\label{table1}
\end{table}

\newpage

\begin{figure}
\epsfxsize=7.2cm
\epsfbox{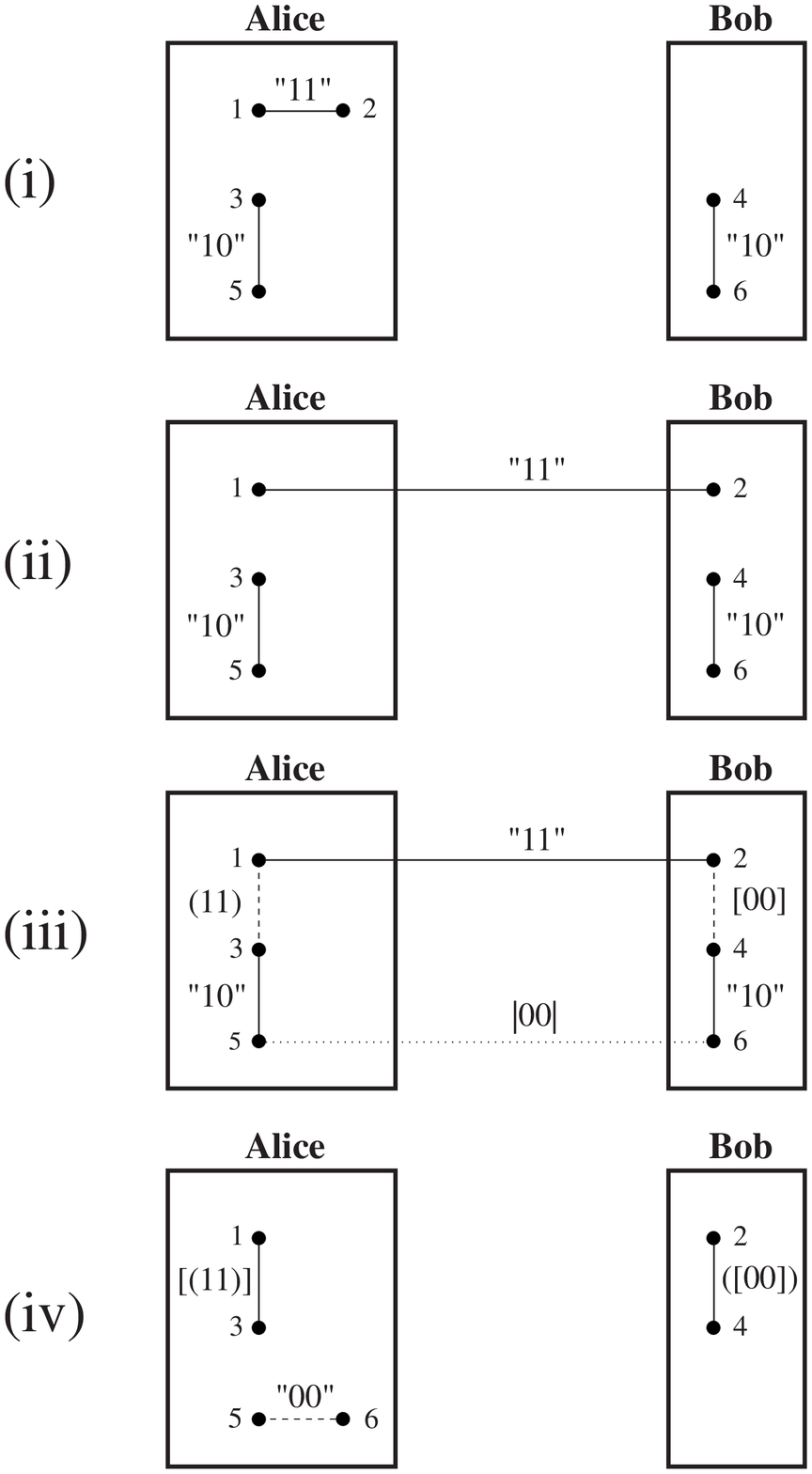}
\end{figure}
\vspace{0.2cm}
\noindent FIG.\ 1. {\small QKD scheme based on entanglement
swapping. The bold lines connect qubits in Bell states, the dashed
lines connect qubits on which a Bell operator measurement is made,
and the pointed lines connect qubits in Bell states induced by
entanglement swapping. ``$00$'' means that the Bell state $\left|
{00} \right\rangle$ is public knowledge, $(00)$ means that it is
only known to Alice, $[00]$ means that it is only known to Bob,
$|00|$ means that it is unknown to all the parts, $[(00)]$ means
that it is only known to Alice and Bob, etc.}

\newpage

\begin{figure}
\epsfxsize=7.2cm
\epsfbox{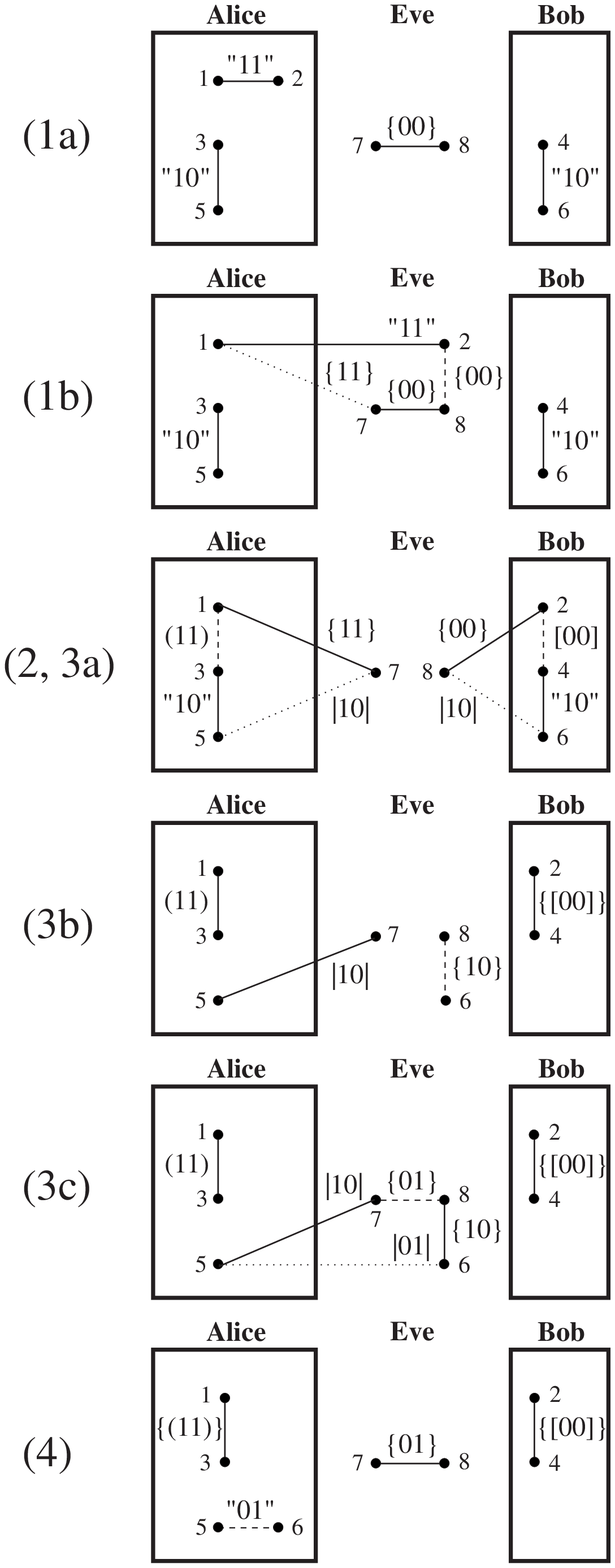}
\end{figure}
\vspace{0.2cm}
\noindent FIG.\ 2. {\small Eve's strategy to obtain Alice's
secret result. $\{00\}$ means that the Bell state $\left| {00}
\right\rangle$ is only known to Eve. The remaining notation is the
same as in Fig.\ 1.}


\begin{thebibliography}{99}
\bibitem{PRL1} Existing schemes for quantum key distribution
can also work with a small numbers of qubits, as long as they are
symmetrised so that in each round of the protocol
Alice and Bob reverse their roles.
\end{thebibliography}
\end{document}